# Stabilizing post-yielding behavior of a stretching dominated lattice structure through microstructural optimization

Mathis Duport[a], Guilhem Martin[a*], Pierre Lhuissier[a], Jean-Jacques Blandin[a], Frédéric Prima[b], Rémy Dendievel[a]

[a] Université Grenoble Alpes, CNRS UMR 5266, Grenoble INP, Laboratoire SIMaP, Grenoble, 38000, France

[b] PSL Research University, Chimie ParisTech, Institut de Recherche de Chimie Paris, CNRS UMR 8247, 75005 PARIS, France

*corresponding author : guilhem.martin@simap.grenoble-inp.fr

## Abstract

We investigate the effect of the microstructure on the mechanical response of lattice structures made of a defect tolerant Ti-14 Mo binary alloy fabricated by Electron Powder Bed Fusion. Some of the lattice structures were subjected to different heat treatments to compare different microstructures: a two-phase α+β microstructure inherited from the fabrication (as-built), a full β-metastable microstructure, and a β-metastable microstructure strengthened via nanoprecipitation of the $\omega_{iso}$ phase. We demonstrate that the macroscopic mechanical behavior of one of the most popular stretching-dominated lattice structure, namely the octet-truss, can be tailored through microstructural optimization by increasing the tolerance to plastic buckling of the constitutive struts. Defect-tolerant materials pave the way towards lattice structures fabricated by additive manufacturing with optimized energy absorption performances.

**Keywords:** Ti-alloy ; Additive Manufacturing ; Microstructure; Lattice structures; Mechanical behavior; Energy absorption

Lattice structures are a class of cellular structures made of a periodic assembly of struts that can deform either by the bending or stretching of the struts [1,2]. Such structures were found interesting because they are good candidates to fill in gaps in the materials properties space as suggested in [1,3]. Stretching-dominated structures were developed for lightweight structural applications, i.e. for high strength and stiffness at low weight [1,3]. Bending-dominated structures are preferred in applications where energy absorption capacity is of utmost importance such as for crush applications [1,3,4]. Different properties related to the energy absorption capacity can be of interest: absorbed energy or energy efficiency. Based on the schematic shown in **Figure 1** and representing a typical compressive response of a lattice structure, we recall the definitions of the different properties related to the energy absorption capacity. The peak-stress corresponds to the maximum stress before the first localization ($\sigma_{peak}$) and the volumetric energy absorption ($U_V$) is estimated by integrating the area under the stress-strain curve until reaching densification strain ($\varepsilon_D$). There is no consensus on the definition of $\varepsilon_D$. It is defined here as the strain corresponding to $\sigma_{peak}$. The absorption energy efficiency is defined as follows: $E(\%) = U_V/U_{id}$ with $U_{id}$ being the volumetric energy absorbed in presence of perfect plateau stress while approaching the peak stress, see **Figure 1**. Finally, the specific energy absorption (*SEA* in J/g) is calculated as follows: $SEA = U_V/\rho_{lattice}$ with $\rho_{lattice}$ the density of the lattice structure. Energy efficiency is particularly relevant when the energy must be absorbed while keeping the maximum transmitted stress (peak stress) below a given threshold stress that would cause injury or damage. For instance, mechanical responses exhibiting a post-yielding strengthening or softening can be considered as less efficient in comparison with a perfectly plastic response.

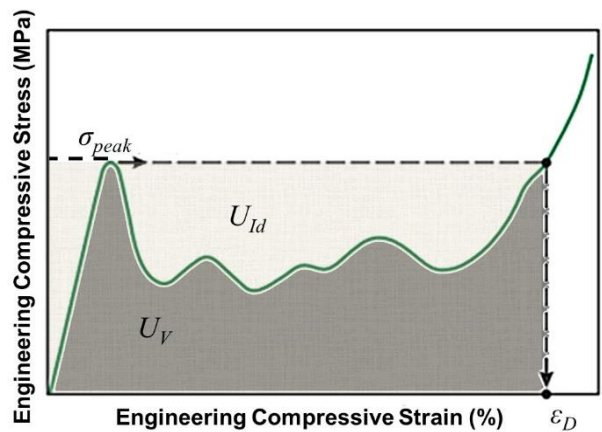

*Figure 1. Schematic of the mechanical response of a lattice structure including the definition of different mechanical properties allowing to quantify the energy absorption performances.*

The advent of Powder Bed Fusion (PBF) additive manufacturing (AM) has made easier the fabrication of complex geometries such as lattice structures, see e.g. [3–6]. Absorbed energy or energy efficiency are related to the post-yielding behavior. This behavior depends of course both on the architecture (e.g. stretching-dominated or bending-dominated) and on the constitutive material, but also on microstructure and defects inherited from AM. For example, pores and surface defects such as notch-like defects can lead to a reduction of ductility of the constitutive material [7] and thus to the early failure of lattice structures. The latter is particularly true when the constitutive material shows a limited ductility and a lack of work-hardening capacity. Using post-fabrication surface treatments can be considered to partially address those limitations [8–11]. In particular, it is difficult to homogeneously apply such treatments to lattice structures. That is why it is thought that using a defect-tolerant constitutive material would be more efficient because the presence of defects would not necessarily lead to early catastrophic failure [12–14]. Damage-tolerant alloys are often materials exhibiting deformation mechanisms such as TRIP (Transformation Induced Plasticity) or TWIP (TWinning Induced Plasticity) because their excellent work-hardening capacity drastically improve their resistance to strain localization. Some authors have recently reported the successful fabrication of TRIP [12,15–17] or TWIP [13,14,18] steels by AM. For lightweight applications, Ti-alloys are good candidates due their good specific strength. However, the heritage Ti-alloys such as the Ti-6Al-4V or the Ti-5Al-5V-5Mo-3Cr alloy are known to suffer from a limited ductility and a poor work-hardening capacity. To overcome such limitations, β-metastable Ti-alloys showing good strength-ductility trade-off with large work-hardening rates thanks to the TRIP/TWIP effects have been developed, see e.g. [19–21]. Binary Ti-Mo alloys have been proven promising candidates with the TRIP/TWIP Ti-12Mo [20,22–24] and the TWIP Ti-15Mo alloy [25].

Tuning the mechanical response of lattice structures by playing with the architecture [26,27] or by varying the relative density [1,2] has been the focus of many research studies, see e.g. [3–5]. However, tailoring the mechanical response of such structures by taking advantage of the metallurgy of the constitutive materials has received little interest and most studies were focused on lattice structures made of Inconel 718, see e.g. [28–30]. Here, we demonstrate that the mechanical behavior of one of the most popular stretching-dominated lattice structures, namely the octet-truss, can be tailored through microstructural optimization of a defect-tolerant binary Ti-14Mo alloy to accommodate plastic buckling of the constitutive struts.

Pre-alloyed powder of binary Ti-14Mo (wt%) alloy was loaded into an E-PBF ARCAM A1 machine operating under vacuum ($2.10^{-3}$ mbar) and at an accelerating voltage of 60 kV. The processing conditions were preliminary optimized ($T_{process}$ = 660°C ± 30°C, $P$ =270 W, *Speed Function* = 130) so that samples with a relative density exceeding 99.90% were fabricated. Bar samples with dimensions 10×10×60 mm$^3$ were produced vertically using a snake-like scanning strategy (hatch spacing = 0.1 mm) with a 90° rotation of the scanning pattern at each layer. Octet-truss lattice structures consisting of 3×3×3 unit cells with struts of 1 mm diameter were also fabricated. Bulk samples and lattice structures were subjected to different heat treatments to produce homogeneous microstructures in an Ar-atmosphere to limit oxygen contamination. Annealing above the β-transus at 900°C for 30 minutes was followed by oil quenching (OQ) to achieve a full β-metastable microstructure [22]. Some samples with a full β-metastable microstructure were also subjected to additional ageing at 150°C for 8h to trigger nanoprecipitation of the $\omega_{iso}$ phase as proposed in [22,23].

Samples were prepared for optical microscopy: grinding down to 2400 using SiC abrasive papers and polishing with a solution consisting of 90% colloidal silica suspension (0.04 μm particle size) and 10% $H_2O_2$. The polished cross-sections were etched with Kroll's reagent to reveal the microstructures. Optical micrographs were taken with an Olympus DSX 510 microscope. SEM images were acquired with a ZEISS GEMINI SEM500 with an operating voltage of 15 kV. For EBSD analysis, additional surface preparation was performed using the VIBROMET from Buehler for 12h using a mixture of a colloidal silica suspension and $H_2O_2$. Electron Back-Scattered Diffraction (EBSD) maps were collected using a step size of 5 μm and post-treated using the OIM Analysis software. X-ray computed tomography was employed to assess the material integrity and to determine the volume of the as-fabricated lattice structure (voxel size 25 μm) using a EasyTom Nano-XL tomograph from RX solutions. Microhardness measurements were performed using a Wilson Tukon 1212 Vickers Tester with a load of 1 kg (HV1). Dog-bone specimens with a gauge length of 15 mm were extracted from the bar by electron discharge machining to determine the tensile response of the various microstructures using a DY35 testing machine equipped with a 20 kN load cell at a strain rate of $10^{-3}$ $s^{-1}$. Three samples for each condition were tested to ensure statistical consistency. Uniaxial compression tests using an MTS 810 equipped with a 100 kN load cell were carried out at $10^{-3}$ $s^{-1}$ to characterize the energy absorption performances of the octet-truss structures subjected to different heat treatments. Digital Image Correlation (DIC) was employed for strain measurements throughout the tests. 6 Mpixels cameras with 50 mm lenses and GOM Correlate Software were used.

**Figure 2** gives an overview of the microstructures investigated in this work. The as-built microstructure is a two phase α+β microstructure as revealed by the SEM-BSE image and EBSD phase map shown respectively in **Figure 2a** and **Figure 2b**. Due to preheating during E-PBF, the as-built microstructure varies along the building direction with a gradient in α-laths size. Here, we provide a typical micrograph taken in the region corresponding to the gauge length of an as-built specimen. β-grains show a typical columnar morphology with an average width between 50 and 80 μm (**Figure 2a**). Such columnar grains result from epitaxial growth from the underlying layers. Two populations of α-laths can be distinguished based on their size. Coarse α-laths are thought to have nucleated and grown in the intergranular regions as well as within the β-grains at a temperature close to the preheating temperature during E-PBF (**Figure 2b**). Intragranular fine α-laths revealed using high magnification images as typically shown in **Figure 2c** have likely formed during the final cooling stage once the build was completed. A full metastable β-microstructure results from annealing at 900°C/30 min followed by OQ (**Figure 2d**). The columnar grain morphology was preserved but their size has increased in comparison with the ones observed in the as-built conditions (average width between 100 and 200 μm). Finally, the full β-microstructure was subjected to additional low temperature ageing at 150°C/8h to strengthen the material via nanoprecipitation of the $\omega_{iso}$ phase. At the grain scale, no difference with the β-metastable microstructure can be evidenced, see comparison between **Figure 2d** and **Figure 2e**. However, the hardness has substantially increased from 280 ± 5 HV1 in the β-metastable microstructure to 350 ± 5 HV1 after ageing at 150°C/8h because of the nanoprecipitation of the $\omega_{iso}$, as documented in [22,23].

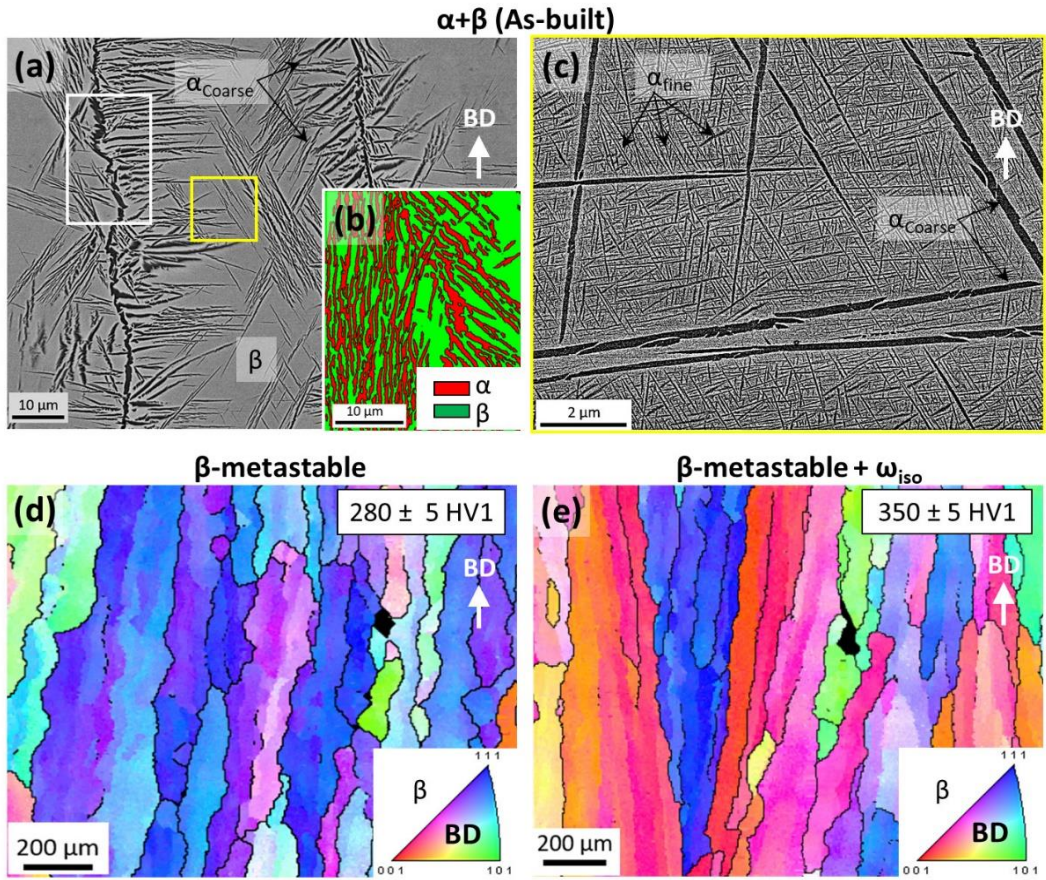

*Figure 2. Overview of the as-built microstructure before and after heat treatment: (a) Low magnification SEM-BSE micrograph showing intergranular and intragranular coarse α-laths embedded in a matrix β; (b) EBSD-phase map confirming the presence of the α and β phase; (c) High magnification SEM-BSE micrograph showing the presence of nanoscale fine α-laths. EBSD-IPF maps revealing the grain structure of (d) the β-metastable microstructure after 900°C/30 min + OQ, and (e) the β-metastable microstructure strengthened via nanoprecipitation of the $\omega_{iso}$ phase (900°C/30 min + OQ + 150°C/8h).*

The tensile response associated with the different microstructures shown in **Figure 2** is displayed in **Figure 3a-b**. The mechanical properties of the different microstructures are summarized in **Table 1**. The two-phase α+β microstructure is the one showing the largest yield strength (nearly 850 MPa) but the smallest elongation to failure (≈10%) due to a limited work-hardening (absence of TWIP effect). On the opposite, the β-metastable microstructure has a yield strength of 450 MPa but an extended ductility (elongation to failure exceeding 43%) thanks to an outstanding work-hardening capacity (**Figure 3a**). Such properties are consistent with the ones reported in the literature for the Ti-12Mo [20,22,24] and Ti-15Mo [25] fabricated by more conventional processing routes. Here, the high work-hardening capacity is essentially due to a TWIP-effect as illustrated in **Figure 3c-d**. The latter microstructure can thus be qualified as defect-tolerant. The β-metastable microstructure strengthened via nanoprecipitation of the $\omega_{iso}$ phase shows an increased yield strength (640 MPa) in comparison with the β-metastable microstructure (450 MPa) while maintaining a relatively good ductility because the TWIP effect is still activated.

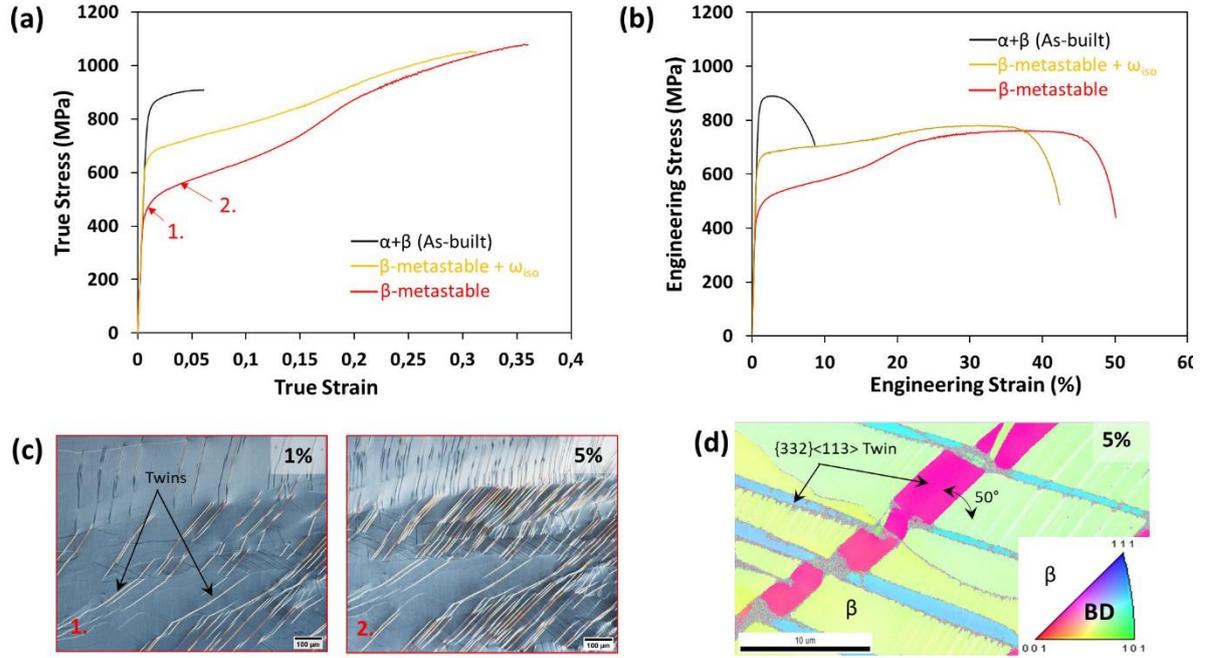

*Figure 3. (a) True stress-strain curves and (b) Engineering stress-strain curves for the different microstructures shown in **Figure 2**. (c) Optical micrographs taken after different plastic strain increments in the same region of interest showing the formation of a twinning network. (d) EBSD-IPF map revealing the nature of the twins: {332} <113> with their typical misorientation of 50°.*

|  | $\sigma_{y,0.2\%}$ (MPa) | $\sigma_{UTS}$ (MPa) | $\Delta\sigma = \sigma_{UTS} - \sigma_{y,0.2\%}$ (MPa) | $\varepsilon_u$ | %A (%) |
|---|---|---|---|---|---|
| **Two-phase α+β** As-built | 845 ± 14 | 920 ± 7 | 71 ± 7 | 0.05 ± 0.02 | 10 ± 2 |
| **β-metastable** 900°C/30 min + OQ | 450 ± 2 | 1046 ± 26 | 598 ± 28 | 0.32 ± 0.04 | 43 ± 6 |
| **β-metastable + ω$_{iso}$** 900°C/30 min + OQ + Ageing 150°C/ 8h | 643 ± 5 | 1034 ± 18 | 390 ± 20 | 0.29 ± 0.02 | 39 ± 3 |

*Table 1. Summary of the mechanical properties of the different microstructures. Average values over the three tests performed are given. $\sigma_{y,0.2\%}$, $\sigma_{UTS}$, $\Delta\sigma$ and $\varepsilon_u$ have been extracted from true-stress-strain curves, and %A from engineering-stress-strain curves.*

Examples of octet-truss lattice structures fabricated by E-PBF are shown in **Figure S 1a**. XCT scans of the lattice structures (see example in **Figure S 1b**) allow the material integrity to be assessed but also the density ($\rho_{lattice} = m_{lattice}/V_{bounding\,box}$) and relative density ($\rho_{lattice}/\rho_{Ti14Mo}$, here ≈ 11%) of the fabricated lattices to be estimated. The latter is important if one aims at properly estimating the specific energy absorption (in J/g) using the density of the fabricated lattice ($\rho_{lattice}$). **Figure S 1c** is a cumulative projective view of all slices along the lattice thickness proving the successful fabrication with a nearly internal defect-free lattice structure.

The compressive mechanical response of the octet-truss lattice structure having an as-built microstructure is first compared to the one showing a β-metastable microstructure, see **Figure 4a**. The microstructures in the lattice structure were found to be very similar to the ones shown in **Figure 2** and observed on bulk samples. The mechanical response of the octet-truss lattice structure having a β-metastable microstructure strengthened by nanoprecipitation of ω$_{iso}$ is also given in **Figure 4a**. Different mechanical properties related to the energy absorption capacity and defined in **Figure 1** have been extracted from **Figure 4a**. The absorption energy properties of the octet-truss structure showing different microstructures are given in **Table 2**.

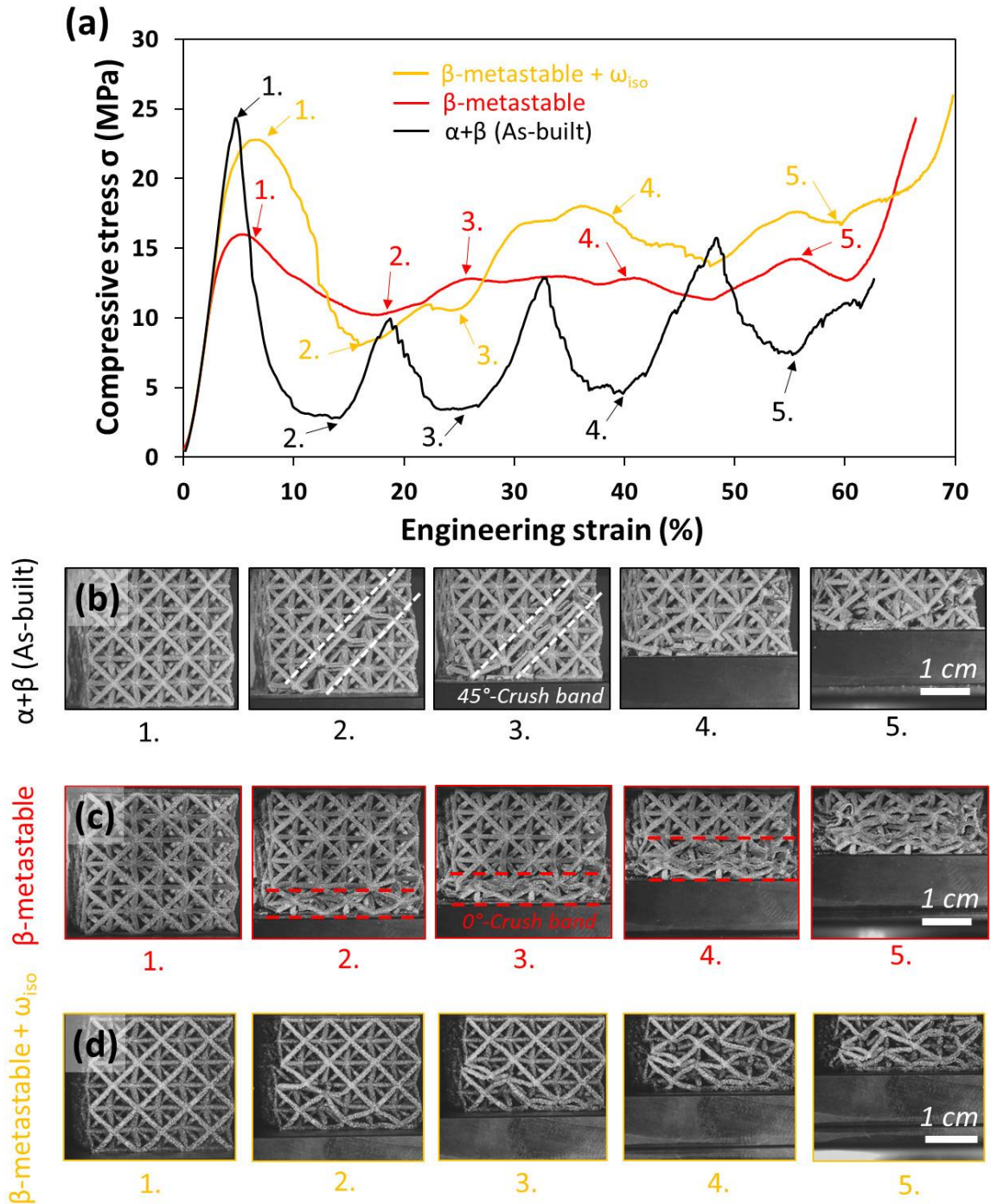

*Figure 4. (a) Compressive engineering stress-strain response of octet-truss lattice structures with an as-built α+β microstructure, a β-metastable microstructure, and with a β-metastable microstructure strengthened by nanoprecipitation of $\omega_{iso}$. Images of the lattice structure taken during mechanical testing of the octet-truss with (b) an as-built α+β microstructure, (c) a β-metastable microstructure, and (d) a β-metastable microstructure strengthened by nanoprecipitation of $\omega_{iso}$.*

The octet-truss structure exhibiting a two-phase α+β microstructure shows an unstable stress-strain response where different regimes can be seen. Elastic regime up to the peak stress is followed by an abrupt drop in stress due to the occurrence of strain localization in a crush band oriented at 45° with respect to the loading direction (**Figure 4b**). The breaking of struts along the crush band is evidenced as illustrated in **Figure 4b**. Then, successive increases and decreases in stress are observed and related to the occurrence of sequential strain localization that goes along with the brittle crush of additional struts, see snapshots at selected strains in **Figure 4b**. The full video is given in **Multimedia 1.**

The octet-truss structure exhibiting a β-metastable microstructure shows a mechanical behavior that contrasts with that of the α+β. Interestingly, the mechanical response of one of the most popular stretching-dominated lattice structure, namely the octet-truss, is tailored through microstructural optimization by stabilizing its post-yielding behavior, the peak stress being immediately followed by plateau stress, see **Figure 4a**. After reaching the peak stress, the abrupt stress decrease is not detected because the struts can sustain large plastic strain before breaking thanks to the activation of the TWIP effect. Most of the struts can be deformed plastically by buckling (see snapshots at selected strains in **Figure 4c**, full video available in **Multimedia 2**). While the stiffness was found nearly unchanged, the strength of the structure (peak strength) was decreased by nearly 30% (24 MPa for the α+β microstructure vs. 16 MPa for the β-metastable microstructure) but the absorption energy efficiency has been significantly increased from about 35% to nearly 80%, see **Table 2**. The recent work of Banait *et al.* [28] is particularly relevant to discuss our results because they investigated the effect of microstructure on the mechanical response of Inconel 718 BCC-lattice structures processed by L-PBF. They show that there is a transition in the mechanical behavior of 3D printed BCC-lattice structures resulting from precipitation-strengthening of the constitutive material upon ageing with a more unstable post-yielding regime. Here, we show using an octet-truss geometry that the post-yielding regime can be stabilized by promoting the formation of a β-metastable microstructure undergoing a TWIP effect instead of the two phase α+β microstructure present in the as-built lattice structure.

|  | $\sigma_{peak}$ (MPa) | $\sigma_{peak}/\sigma_{y,0.2\%}$ | $U_V$ (mJ/mm$^3$) | *SEA* (J/g) | *E* (%) |
|---|---|---|---|---|---|
| **Two-phase α+β** As-built | 24.4 | 0.029 | 5.2 | 9.7 | 35.1 |
| **β-metastable** 900°C/30 min + OQ | 16.0 | 0.036 | 7.8 | 14.6 | 79.7 |
| **β-metastable + ω$_{iso}$** 900°C/30 min + OQ + Ageing 150°C/ 8h | 22.8 | 0.035 | 10.5 | 19.6 | 69.9 |

*Table 2. Energy absorption properties of the octet-truss lattice structure with different microstructures.*

The mechanical response of the octet-truss with a β-metastable microstructure was further tuned by strengthening the matrix via nanoprecipitation of the ω$_{iso}$ phase. An intermediate mechanical behavior (**Figure 4a**) is obtained. Through nanoprecipitation, the strength of the lattice structure exhibiting the α+β microstructure is almost recovered but, at the same time, its sensitivity to strain localization is greatly reduced since struts were still able to sustain large plastic strain (**Figure 4d**) thanks to a preserved TWIP effect. This β-metastable microstructure strengthened with $\omega_{iso}$ precipitates allows to further increase the volumetric absorption: $U_V = 10.5$. mJ/mm$^3$ vs. $U_V = 7.8$ mJ/mm$^3$ for the β-metastable microstructure. However, the post-yielding behaviour is more unstable, therefore the energy absorption efficiency is reduced in comparison with the β-metastable microstructure (70% vs. nearly 80%). Finally, it worth noticing that in the post-yielding regime, the amplitude of stress variations seems related to the work-hardening capacity of the constitutive material. The microstructure showing the poorest work-hardening capacity is the one exhibiting the largest stress variations. On the opposite, the β-metastable microstructure shows smooth variations of stress in the post-yielding regime.

Using defect-tolerant materials to build stretching-dominated lattices structures broadens the range of applications of such materials because their microstructure can be tuned to become tolerant to plastic buckling of the constitutive struts in order to stabilize the post-yielding behavior. Optimizing the microstructure of lattice structures processed by additive manufacturing paves the way towards lattice structures with optimized energy absorption performances and extend the applications of stretching-dominated structures.

## Declaration of competing interest

The authors declare that they have no competing interests.

## Acknowledgements

This work was funded by the Center of Excellence of Multifunctional Architectured Materials "CEMAM" n°AN-10-LABX-44-01. This work has benefited from the facilities available on the characterization platform CMTC of Grenoble INP.

## References

[1] M.F. Ashby, R.F.M. Medalist, MTA 14 (1983) 1755–1769.
[2] M.F. Ashby, Phil. Trans. R. Soc. A. 364 (2006) 15–30.
[3] J. Bauer, L.R. Meza, T.A. Schaedler, R. Schwaiger, X. Zheng, L. Valdevit, Adv. Mater. 29 (2017) 1701850.
[4] M. Pham, C. Liu, I. Todd, J. Lertthanasarn, Nature (2019).
[5] L.-Y. Chen, S.-X. Liang, Y. Liu, L.-C. Zhang, Materials Science and Engineering R 146 (2021) 56.
[6] E. Alabort, Scripta Materialia (2019) 5.
[7] T. Persenot, G. Martin, R. Dendievel, J.Y. Buffiére, E. Maire, Materials Characterization 143 (2018) 82–93.
[8] C. de Formanoir, M. Suard, R. Dendievel, G. Martin, S. Godet, Additive Manufacturing 11 (2016) 71–76.
[9] P. Lhuissier, C. de Formanoir, G. Martin, R. Dendievel, S. Godet, Materials and Design 110 (2016) 485–493.
[10] G. Pyka, G. Kerckhofs, I. Papantoniou, M. Speirs, J. Schrooten, M. Wevers, (2013) 4737–4757.

[11] B.G. Pyka, A. Burakowski, G. Kerckhofs, M. Moesen, S.V. Bael, J. Schrooten, M. Wevers, 14 (2012) 363–370.
[12] J. Günther, F. Brenne, M. Droste, M. Wendler, O. Volkova, H. Biermann, T. Niendorf, Sci Rep 8 (2018) 1298.
[13] M.S. Pham, B. Dovgyy, P.A. Hooper, Materials Science and Engineering: A 704 (2017) 102–111.
[14] F. Kies, P. Köhnen, M.B. Wilms, F. Brasche, K.G. Pradeep, A. Schwedt, S. Richter, A. Weisheit, J.H. Schleifenbaum, C. Haase, Materials & Design 160 (2018) 1250–1264.
[15] M. Droste, J. Günther, D. Kotzem, F. Walther, T. Niendorf, H. Biermann, International Journal of Fatigue 114 (2018) 262–271.
[16] C. Burkhardt, R. Wagner, C. Baumgart, J. Günther, L. Krüger, H. Biermann, Adv. Eng. Mater. 22 (2020) 2000037.
[17] P. Köhnen, C. Haase, J. Bültmann, S. Ziegler, J.H. Schleifenbaum, W. Bleck, Materials & Design 145 (2018) 205–217.
[18] X. Wang, J.A. Muñiz-Lerma, O. Sanchez-Mata, M. Attarian Shandiz, N. Brodusch, R. Gauvin, M. Brochu, Scripta Materialia 163 (2019) 51–56.
[19] Y. Danard, G. Martin, L. Lilensten, F. Sun, A. Seret, R. Poulain, S. Mantri, R. Guillou, D. Thiaudière, I. Freiherr von Thüngen, D. Galy, M. Piellard, N. Bozzolo, R. Banerjee, F. Prima, Materials Science and Engineering: A 819 (2021) 141437.
[20] M. Marteleur, F. Sun, T. Gloriant, P. Vermaut, P.J. Jacques, F. Prima, Scripta Materialia 66 (2012) 749–752.
[21] F. Sun, J.Y. Zhang, M. Marteleur, C. Brozek, E.F. Rauch, M. Veron, P. Vermaut, P.J. Jacques, F. Prima, Scripta Materialia 94 (2015) 17–20.
[22] F. Sun, J.Y. Zhang, P. Vermaut, D. Choudhuri, T. Alam, S.A. Mantri, P. Svec, T. Gloriant, P.J. Jacques, R. Banerjee, F. Prima, Materials Research Letters 5 (2017) 547–553.
[23] S.A. Mantri, F. Sun, D. Choudhuri, T. Alam, B. Gwalani, F. Prima, R. Banerjee, Sci Rep 9 (2019) 1334.
[24] F. Sun, J.Y. Zhang, M. Marteleur, T. Gloriant, P. Vermaut, D. Laillé, P. Castany, C. Curfs, P.J. Jacques, F. Prima, Acta Materialia 61 (2013) 6406–6417.
[25] X. Min, X. Chen, S. Emura, K. Tsuchiya, Scripta Materialia 69 (2013) 393–396.
[26] V.S. Deshpande, M.F. Ashby, N.A. Fleck, Acta Materialia 49 (2001) 1035–1040.
[27] C. Bonatti, D. Mohr, International Journal of Plasticity 92 (2017) 122–147.
[28] S. Banait, X. Jin, M. Campos, M.T. Pérez-Prado, Scripta Materialia 203 (2021) 114075.
[29] B.B. Babamiri, B. Barnes, A. Soltani-Tehrani, N. Shamsaei, K. Hazeli, Additive Manufacturing 46 (2021) 102143.
[30] B.B. Babamiri, J.R. Mayeur, K. Hazeli, Additive Manufacturing 52 (2022) 102618.

**SUPPLEMENTARY MATERIALS**

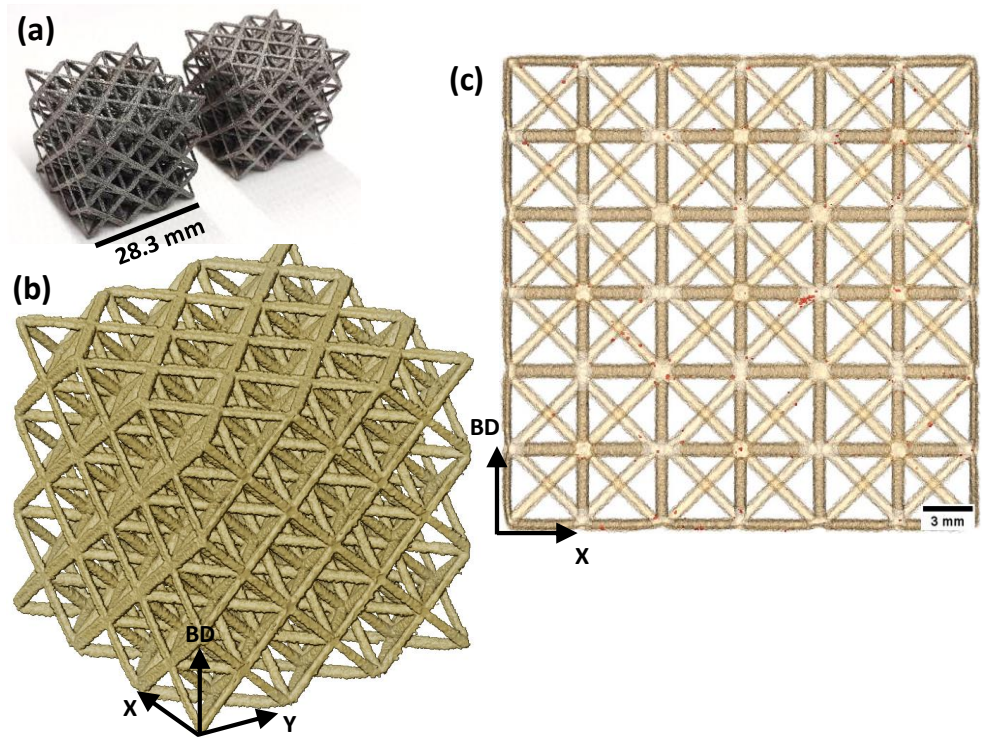

*Figure S 1. (a) Images of as-built octet-truss lattice structures fabricated by E-PBF. (b) 3D rendering of an as-built lattice structure. (c) Axial cumulative projective view of all the slices throughout an as-built lattice structure revealing the presence of very few pores (in red).*

# MULTIMEDIA FILES

**Alpha+Beta_As-Built.mp4** file attached to this submission

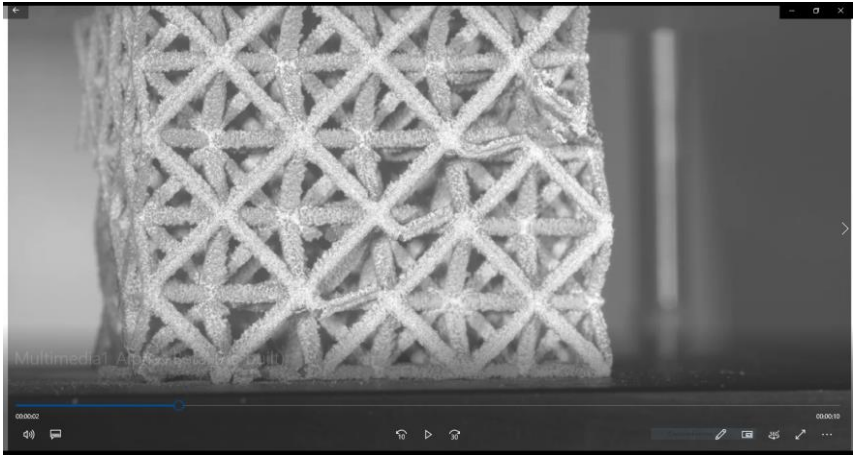

*Multimedia 1: Video recorded during compression of the lattice structure exhibiting a two phase α+β microstructure.*

**Beta_metastable.mp4** file attached to this submission

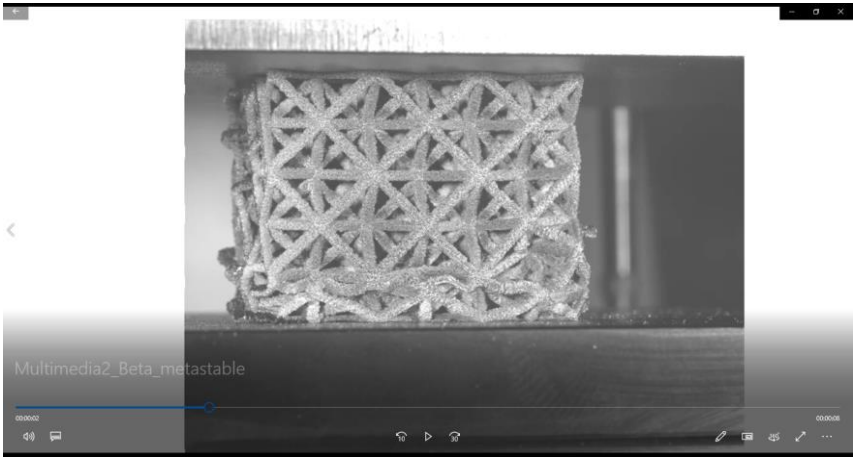

*Multimedia 2: Video recorded during compression of the lattice structure having a β-metastable microstructure (TWIP-effect).*